\documentclass[12pt]{article}
\usepackage{graphicx,amsmath,amssymb,amsthm}
\newcommand{\ssy}[5]{#1  {\it #2}  {\bf #3} (#4) #5\rlap{.}}
\newcommand{\kartih}[3]{\begin{figure}[#1]\begin{center}
\includegraphics[width=\textwidth]{#2.ps}
\end{center}\caption{#3}\end{figure}}

\newcounter{aaa}
\newenvironment{teor}[2][{}]{\begin{trivlist}\refstepcounter{aaa}%
\labelsep=0pt\item[\bfseries \theaaa. #2. ]#1}%
{\end{trivlist}}
\newcommand{\rmd}{\mathrm{d}}

\DeclareMathOperator{\Bd}{Bd}
\DeclareMathAlphabet{\eu}{U}{eur}{m}{n}
\DeclareSymbolFont{iso}{U}{txmia}{m}{it}
\DeclareMathSymbol{\piup}{\mathord}{iso}{"19}
\DeclareMathSymbol{\kappaup}{\mathord}{iso}{"14}
\DeclareMathSymbol{\zetaup}{\mathord}{iso}{"10}
\DeclareMathSymbol{\varthetaup}{\mathord}{iso}{"23}
\DeclareMathSymbol{\omegaup}{\mathord}{iso}{"21}
\DeclareMathSymbol{\varphiup}{\mathord}{iso}{"27}
\DeclareMathSymbol{\chiup}{\mathord}{iso}{"1F}
\DeclareMathSymbol{\nat}{\mathalpha}{iso}{"8E}
\DeclareMathSymbol{\rea}{\mathalpha}{iso}{"92}
\DeclareMathSymbol{\eucl}{\mathalpha}{iso}{"85}
\DeclareMathSymbol{\mink}{\mathalpha}{iso}{"8C}
\DeclareMathSymbol{\cel}{\mathalpha}{iso}{"9A}
\DeclareMathSymbol{\Mf}{\mathalpha}{iso}{"8D}
\DeclareMathSymbol{\okr}{\mathalpha}{iso}{"93}
\newcommand*{\s}{\eu{S}}
\author{S. Krasnikov}
\title{Unconventional string-like singularities in flat spacetime.}
\date{}
\begin{document}
\maketitle
\begin{abstract}
The conical singularity in flat spacetime is mostly known as a model of
the cosmic string or the wedge disclination in solids. Another, equally
important, function is to be a representative of  quasiregular
singularities. From all these points of view it seems interesting to find
out whether there exist other similar singularities. To specify what
``similar" means I introduce the notion of the string-like singularity,
which is, roughly speaking, an absolutely mild singularity concentrated
on a curve or on a 2-surface $\s$ (depending on whether the space is
three- or four-dimensional). A few such singularities are already known:
the aforementioned conical singularity, its two Lorentzian versions, the
``spinning string", the ``screw dislocation", and Tod's spacetime. In all
these spacetimes $\s$ is a straight line (or a plane) and one may wonder
if this is an inherent property of the string-like singularities. The aim
of this paper is to construct string-like singularities with less trivial
$\s$. These include flat spacetimes in which $\s$ is a spiral, or even a
loop. If such singularities exist in nature (in particular, as an
approximation to gravitational field of strings) their cosmological and
astrophysical manifestations must differ drastically from those of the
conventional cosmic strings. Likewise, being realized as topological
defects in crystals such loops and spirals will probably also have rather
unusual properties.
\end{abstract}
\newpage
\section{Introduction}
Consider the spacetime $\eu{M}_{1}$:
\begin{equation}\label{eq:cone}
\begin{split}
\rmd s^2= -\rmd t^2 + \rmd z^2 + \rmd \rho^2 + \rho^2\rmd \phi^2,
\\
  t,z\in \rea,\quad \rho>0,\quad \phi=\phi+a
\end{split}
\end{equation}
(the last formula means that $\phi$ parameterizes the circle and is
defined modulo $ a$ only). When $a=2\pi$ the spacetime  is merely the
Minkowski space $\mink^4$ in which the timelike plane $\rho=0$ is
deleted. However, if $a$ takes any other non-zero value, $\eu{M}_{1}$
becomes a quite non-trivial spacetime often referred to as a `straight
cosmic string'. $\eu{M}_{1}$ is evidently singular and it is the
singularities of this type that are discussed in this paper under the
general name of `string-like singularities' (the words `this type'  need
some elucidation of course and it will be given in due course). These
singularities are important in many ways:

\emph{Cosmology.} It is widely believed that the phase transitions in the
early universe could result in formation of cosmic strings --- infinitely
long and at the same time very thin solutions of the combined system of
Einstein, Higgs and gauge field equations (see, e.~g., \cite{obz} for
reviews and references).  No exact solutions of that system are known,
but Vilenkin \cite{Vilenkin81}  argued that in  some approximation a
universe with a static cylindrically-symmetric cosmic string, when in
addition the metric is invariant under boosts along the string, is
described by a spacetime $U$ that coincides with $\eu{M}_{1}$ at
sufficiently large $\rho$: $\rho>\rho_0$. In this sense $\eu{M}_{1}$ is
an approximation --- useful when $\rho_0$ can be neglected --- of the
singularity-free string spacetime $U$; the latter can be called a
`thickening' of $\eu{M}_{1}$.

The singularities considered in this paper are defined by two properties:
they (in the sense yet to be elaborated) are surfaces of co-dimension two
and the spacetimes harboring them are regular (flat in most cases). So,
they can be regarded as generalization of $\eu{M}_{1}$ to less symmetric
case (each of them must of course lack some of the symmetries mentioned
above, e.~g.\ the `spinning string' $\eu{M}^-_{4}$ is not
boost-invariant, while all curved singularities are not
cylindrically-symmetric) and that is why they are called string-like.

All timelike singularities discussed below admit thickening, but it may
happen that for some particular type  the thickenings cannot describe
cosmic strings, because, say, the properties of the matter required (by
the Einstein equations) for their existence are unrealistic. With this
reservation, however, every new class of string-like singularities must
be of great importance to the cosmic string theory, being an
approximation to the gravitational field of (closed, curved, accelerated,
etc.) strings.

\emph{Solid state physics.} The spacetime~\eqref{eq:cone} bears much
resemblance to what is called wedge disclination in condensed matter
physics (see, e.~g., \cite{ChLu}). Another string-like singularity is
similar to screw dislocation (see section~\ref{sec:spstr}). Though the
analogy between the spacetime singularities and the defects from the
theory of elasticity is delusive sometimes (see section~\ref{sec:mixed}
and Appendix~\ref{A:PS}) it can be stretched further
--- Puntigam and
 Soleng used the Volterra construction to classify the string-like
 singularities \cite{PS}. Which suggests that, vice versa, the properties
 of string-like
 singularities might be important in condensed matter
physics.

\emph{Relativity.} What makes the string-like
 singularities especially interesting is, in my view, their relation to
 the most fundamental problems of general relativity. Whether so mild
 singularities exist in nature is, in a sense, a more important question
 than, say, whether there is a singularity inside the Schwarzschild
 horizon. Indeed the singularities in discussion satisfy (see
 definition~\ref{def:str}) the condition --- let us call it \emph{absolute
 mildness} --- that there be a finite open covering of the spacetime $M$
\begin{equation}\label{eq:cov}
M=\bigcup_{i=1,\dots m} U_i
\end{equation}
such that every $U_i$ can be extended to a non-singular spacetime $M_i$
[that is, there are isometries $\omegaup_i$ mapping $U_i$ to
$\omegaup_i(U_i)\subset M_i$].  In fact, the absolutely mild
singularities are --- at least in flat spacetimes
--- a subclass of quasiregular ones
\cite{quasireg, Clarke} the difference being essential only in exotic
situations, when infinitely many quasiregular singularities accumulate in
a spacetime (see Appendix B of \cite{quasireg}). The peculiarity of these
types of singularities is that however close one of them is approached
the geometry remains perfectly nice. This makes their presence in
relativity ruinous for its predictive force: even if a spacetime is
initially globally hyperbolic its evolution  cannot be predicted from the
Cauchy surface, because at any moment a singularity (say, a `branching'
singularity discussed in section~\ref{sec:free}) can form, nullifying all
our predictions. At the same time it is absolutely unclear how to exclude
such singularities from the theory. Unless forbidden by some \emph{ad
hoc} global postulate the same branching singularity would apparently
present in any geometrical theory regardless of the dynamical equations
for the geometry, its relation to the matter source, or the properties of
that source.
\begin{teor}{Remark} The seriousness of the problem is often underestimated.
For example, in their pioneering paper on quasiregular singularities
\cite{quasireg} Ellis and Schmidt speaking through Salviati
say: ``We know lots of examples of quasiregular singularities, all
constructed by cutting and gluing together decent space-times; and
because of this construction, we know that these examples are not
physically relevant." The argument is emphatically untenable: \emph{any}
spacetime can be constructed by cutting and gluing together some other
decent spacetimes and \emph{any} of them can be constructed otherwise.
The spacetimes with the singularities in discussion are absolutely no
different in this respect from the others. Correspondingly, no reasons
are seen to regard $\eu{M}_{1}$ and suchlike less physically relevant
than any other spacetime.
\end{teor}

To sum up, there are many reasons for studying string-like singularities
and, in particular, those occurring in flat spacetimes (supposedly they
are the simplest). The first question that one may ask is: What form do
they have? So far only a few such singularities have been considered in
the literature and all of them (with a possible exception for the
`branching disclination', discussed in section~\ref{sec:free}) have very
dull form: they are flat surfaces of co-dimension 2. In other words, they
(or rather their thickenings) correspond to straight strings moving at
constant velocities. The main aim of the present paper is to provide
examples of flat spacetimes with \emph{different} string-like
singularities including those corresponding to curved
--- and even closed --- strings and strings moving with acceleration.
This will be done in section~\ref{Sec:non-ES} after some general
consideration in section~\ref{sec:gen}, where I define string-like
singularities and (roughly) classify  them.

\section{General consideration} \label{sec:gen}
\subsection{String-like singularities}\label{sec:2.1}
In trying to build a curved or otherwise unusual string-like singularity
one immediately comes up against the problem of definition. It is
customary, for example, to refer to the  singularity in the spacetime
[from now on the word \emph{spacetime} stands for smooth connected
(pseudo-)Riemannian manifold]
\begin{equation}\label{eq:cone_Eu}
\eu{M}_{1'}^+\colon\qquad\rmd s^2= \rmd z^2 + \rmd \rho^2 + \rho^2\rmd
\phi^2,\qquad
\rho>0,\quad\phi=\phi+ a,\quad a\neq 0, 2\pi
\end{equation}
as to the `straight line' while the singularity in $\eu{M}_{1}$ is, in
these terms, a `plane' or `a straight line at rest'. But what
\emph{exactly} is meant by that? The metric, and hence the spacetime,
cannot be extended to the $z$-axis, straight or not. But if the $z$-axis
is missing\footnote{One can try to retain it in the spacetime by
developing `distributional geometry', see
\cite{SteinVick} and references therein.}, then just what is straight or bent? In
fact, this naive question has no good answer at present being a
particular case of a notoriously hard problem of assigning in a natural
way a topology (never mind geometry) to singularities \cite{HawEl}.
Fortunately, when the condition~\eqref{eq:cov} is satisfied it is
possible to give at least a
\emph{tolerable} working definition to the relevant entity.

Consider to this end  the set $\Gamma$ of geodesics $\gamma(\tau)\colon\
[-1,0)\to M$ which cannot be extended to the zero value of the affine
parameter $\tau$. Denote, further, by $\Gamma_i$ the subset of $\Gamma$
which consists of the geodesics lying, at least when $|\tau|$ is
sufficiently small, in $U_i$ [here $U_i$ is a member of the covering
\eqref{eq:cov}]. Below we are only interested in spacetimes and coverings
such that
\begin{equation}\label{uslnagam}
\Gamma=\bigcup_{i=1,\dots m} \Gamma_i
\end{equation}
(this does not follow automatically from \eqref{eq:cov} as can be seen by
example of Misner's space). Though the geodesics $\gamma\in\Gamma_i$ do
not have the endpoints $\gamma(0)$, their images $\omegaup_i\circ\gamma$
in $M_i$ do. We shall denote such endpoints by $\eu{s}$ with
corresponding indices:
\[
\eu{s}_{\gamma,i}=\omegaup_i\circ\gamma(0).
\]
Since  $\gamma$ may lie in more than one $U_i$, it may happen that two
different points $\eu{s},\eu{s}'\in\bigcup_i M_i $ are generated by the
same geodesic. We shall write $\eu{s}_1\sim\eu{s}_2$ in such cases :
\[
\eu{s}\sim\eu{s}'\quad\Longleftrightarrow\quad
 \exists\,i_1,i_2,\gamma\colon\quad
\eu{s}=\eu{s}_{\gamma,i_1},\ \eu{s}'=\eu{s}_{\gamma,i_2}.
\]
It would be natural to identify $\eu{s}$ and $\eu{s}'$ and to associate
the singularity with the quotient of  $\s_\Gamma=\bigcup
\eu{s}_{\gamma,i}$ over
 $\sim$, but unfortunately in the general case $\sim$ is not an equivalence
relation. Therefore, we introduce the equivalence relation $\eqsim$ as
the transitive closure of $\sim$:
\[
\eu{s}_1\eqsim\eu{s}_2\quad\Longleftrightarrow\quad
\eu{s}_1\thicksim\eu{s}_{k_1}\thicksim\eu{s}_{k_2}\dots
\thicksim\eu{s}_{k_m}\thicksim\eu{s}_2
\]
and use  $\eqsim$ in checking whether a candidate set represents the
entire singularity.
\begin{teor}{Definition}\label{def:str} Let conditions \eqref{eq:cov} and
\eqref{uslnagam} hold  in a spacetime $M$. Then a set
$\s\subset\s_\Gamma$ is said to \emph{represent} the singularity of  $M$
if for any $\eu{s}\in\s_\Gamma$ there is $\eu{s}'\in\s$ such that
$\eu{s}'\eqsim\eu{s}$.
\end{teor}
$\s$ is considered as  subspace of   $\bigcup M_i$ (not just a set of
points) and correspondingly it can be straight, or curved, timelike or
not, etc. So, the definition seems to capture the idea of a singularity
being of a particular form.
\begin{teor}{Remark} The price to be paid is some arbitrariness. First,
depending on the choice of $U_i$ the same singularity can be represented
by different sets. Furthermore, one can argue that geometrically it would
be more consistent to consider the singularity itself, defined, say, as
${\s}_\Gamma/\eqsim$, rather than a set representing it. On the other
hand, in considering strings, i.~e.\ thickenings of the singularities,
$\s$ seems to be more adequate. The difference between the two objects is
exemplified by the `spinning string', see section~\ref{sec:spstr}. $\s$
in that case is a plane, while ${\s}_\Gamma/\eqsim$ is
\emph{a cylinder}.
\end{teor}
Now we can at last delineate our subject more specifically.
\begin{teor}{Definition} A singularity is
\emph{string-like} if it can be represented by a surface of co-dimension
two.
\end{teor}
\begin{teor}{Notation}
In what follows \emph{three-}dimensional spacetimes with string-like
 --- i.~e.\
represented by  \emph{curves} in this occasion --- singularities are
denoted $\eu{M}_{1'}$, $\eu{M}_{2'}$, etc. Some of them differ only in
the signature in the sense that they are obtained by the same, explicitly
prescribed, manipulations applied either to the Euclidean space $\eucl^3$
or to the Minkowski space $\mink^3$ (by the `same' manipulations I mean
that their verbal descriptions becomes the same after the words $z$-axis
and $t$-axis are interchanged; hence the notation $\vartheta$ in the
figures
--- it stands for `$z$ or $t$'). To such spacetimes the same numbers will
be given and they will be denoted by $\eu{M}_{k'}^+$ and
$\eu{M}_{k'}^-$, correspondingly. The
\emph{four-}dimensional spacetimes will be denoted similarly but without
primes: $\eu{M}_{1}$, $\eu{M}_{2}$, etc. And the correspondence rule is:
 given an
$\eu{M}_{k'}^{(\pm)}$ one obtains its four-dimensional version (i.~e.\
$\eu{M}_{k}^{(\pm)}$)  by simply multiplying the former by  the relevant
axis. For example,
\[
\eu{M}_{1}=
\eu{M}_{1'}^-\times\eucl^1\quad\text{or}\quad
 \eu{M}_{4}^{+}=\eu{M}_{4'}^+\times\mink^1, \quad \text{etc.}
\]
\end{teor}

The first family of string-like singularities was constructed by Ellis
and Schmidt \cite{quasireg} who produced them from the  sets of fixed
points $\eu F$ of discrete isometries $\zetaup$ acting on the Minkowski
space, see appendix~\ref{A:ES}, much as one obtains the usual
two-dimensional cone by identifying the points on a plane related by a
rotation by some fixed angle. The point is that the spacetime
$(\mink^4-\eu F)/\zetaup$ is singular (the geodesics which in the
Minkowski space terminated at $\eu F$ now have no endpoints) and its
singularity (which is irremovable due to the nature of $\eu F$, cf.
section 5.8 in \cite{HawEl}) is represented by $\eu F$. Three $\zetaup$
were considered in
\cite{quasireg} ---  rotation, boost, and boost+rotation --- to obtain in
each case a string-like singularity  represented by a plane. So, it might
have appeared that
\begin{enumerate}\renewcommand{\theenumi}{(\roman{enumi})}
  \item the problem of determining all the elementary string-like
singularities in flat space is essentially equivalent to finding all the
discrete subgroups of the Lorentz group which have two-dimensional
surfaces  $\eu F$ as their sets of the fixed points
\cite{quasireg} and
  \item all such singularities are straight, i.~e.,   $\eu
  F$ are planes \cite{Vickers}.
\end{enumerate}
 In fact, however, neither is true.
Counterexamples to (ii) are built in the next section (an obvious one is
the double covering of $\eucl^3-\eu F$, where $\eu F$ is an arbitrary
curve) and that (i) is not the case is seen from the fact that even the
spacetime
\eqref{eq:cone} with $a>1$ cannot be obtained in that manner (instead of
$\mink^4$ one could have started from a covering of $\mink^4-\eu F$ in
this case, but the relevant isometries have no fixed points).

The reasons why  the requirements to the isometries can be weakened will
become evident from examples in section~\ref{sec:spstr} (roughly speaking
one can produce the desired singularities from the discontinuing set of
an isometry rather from its set of the fixed points), to which we shall
turn after introducing (or, rather, formalizing
--- it is well known and widely used) a more visual method of
constructing spacetimes.

\subsection{Cut--and--paste surgery}\label{sec:surg}

Given $V$ is an open subset of a spacetime one can construct a new
spacetime $W$ in the following way. Pick a pair $p_{1,2}$ of different
points in the boundary $\mathcal B\equiv\Bd V$ and let $O_{1,2}$ be
disjoint neighbourhoods of $p_{1,2}$. Either of the neighbourhoods can be
split into three disjoint sets
\[
O_{j\vee}=O_j\cap V,\qquad \mathcal B_j=O_j\cap\mathcal B ,\qquad
O_{j\wedge}=O_j - B_j-O_{j\vee},\qquad j=1,2
\]
see figure~\ref{fig:gl}.
\kartih{tb}{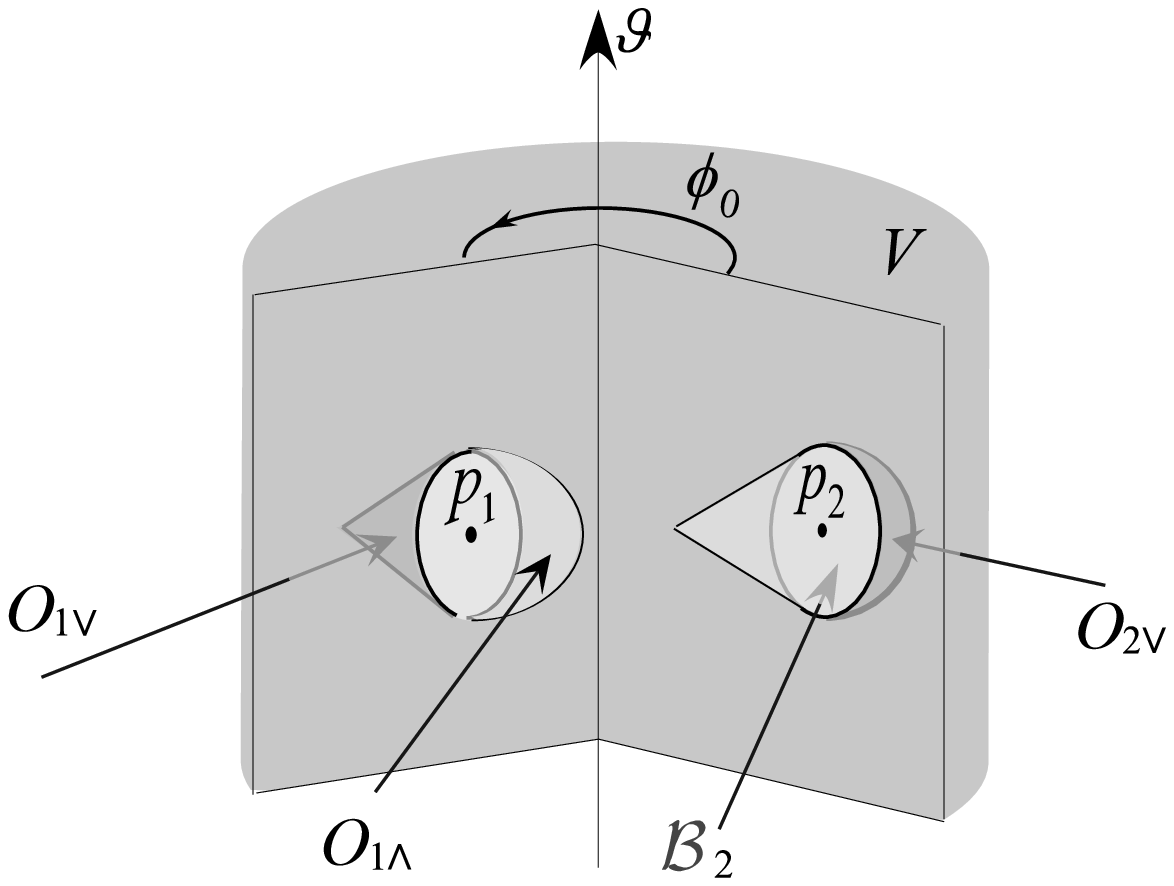}{\label{fig:gl} The tear-shaped regions are $O_1$ and
$O_2$. If $O_1$, $O_2$ are chosen otherwise (namely, to be the sectors
$-\delta<\phi<\delta$ and $\phi_0-\delta<\phi<\phi_0+\delta$), then
 $\mathcal B_1$ and $\mathcal B_2$ are half-planes bounded by the
 $\vartheta$-axis. Gluing them with an appropriate isometry one
obtains all (but $\eu{M}_{3}$ and $\eu{M}_{5}$) singular spacetimes
listed in this section and
Appendix~\ref{A:ES}.}%
Suppose now there is an isometry $\kappaup\colon\ O_1\to O_2$ such that
\[
\kappaup(O_{1\vee})=O_{2\wedge},\qquad\kappaup(\mathcal
B_1)=\kappaup(\mathcal B_2) ,\qquad\kappaup(O_{1\wedge})=O_{2\vee}.
\]
Then an  equivalence relation $\approx$ can be defined
\[
p\approx q\quad\Leftrightarrow\quad
 p=q,\text{ or }\kappaup(q),\text{ or }\kappaup^{-1}(q)
\]
and the spacetime $W$ constructed by identifying equivalent points:
\begin{gather*}
W= (V\cup O_1\cup O_2)/\approx\\ \hspace{10em}=(V\cup \mathcal B_1\cup
\mathcal B_2)/\approx.
\end{gather*}
Pictorially speaking $W$ is obtained by  first attaching to  $V$ two
parts of its boundary, $\mathcal B_1$ and $\mathcal B_2$, and by
 then gluing these parts together  (remarkably, as long as $\mathcal B_1$
 and $\mathcal B_2$ are kept fixed the choice of the
points $p_{1,2}$ and their neighbourhoods $O_{1,2}$ does not affect the
result, they only are needed to guarantee the smoothness of $W$). The
isometry $\kappaup$ has to be explicitly pointed out sometimes because
there may be more than one way of gluing $\mathcal B_1$ to $\mathcal
B_2$.

\begin{teor}{Example}\label{ex:con}
Take $V$ to
be the region $0<\hat\phi<a$ in
\begin{equation}\label{eq:Mcov}
\hat M\colon\qquad\rmd s^2= \rmd \hat z^2 + \rmd \hat \rho^2 + \hat
\rho^2\rmd
\hat \phi^2,\qquad
\hat z,\hat \phi\in \rea^1,\quad \hat \rho> 0
\end{equation}
and glue its boundaries $\mathcal B_1$ and $\mathcal B_2$ (i.~e.\ the
surfaces $\hat\phi=0$ and $\hat\phi=a$, respectively) with the
translation $\kappaup\colon\
\hat\phi\mapsto\hat\phi+a$ (in section~\ref{sec:spstr} we shall glue the
same surfaces with other isometries as well). The result is the conical
space $\eu{M}^+_{1'}$, see \eqref{eq:cone_Eu}.
\end{teor}
\begin{teor}{Remark} Of course the spacetime $W$ can be described as well
in terms of quotient spaces --- it is, for instance, a quotient of its
universal covering  $\widetilde W$. Vice versa, given  $\widetilde W$ we
can build $W$ by cutting a fundamental region from $\widetilde W$ and
gluing appropriately its boundaries. So, it is just a matter of
convenience, which language to use.  In particular, the Ellis--Schmidt
singularities, see appendix~\ref{A:ES},  are also easily constructed by
cut--and--paste surgery.
\end{teor}
Below we shall construct spacetimes much like in example~\ref{ex:con}.
Denote by $\Mf^n$ the Minkowski or Euclidean $n$-dimensional space
\[
\Mf^n=\mink^n,\eucl^n\qquad n=3,4
\]
(which exactly will be indicated explicitly, when important). Pick an
$(n-1)$-dimensional simply connected surface $H\subset \Mf^n$ such that
\begin{enumerate}
\renewcommand{\theenumi}{\roman{enumi})}
  \item $H$ is  invariant under an isometry
$\kappaup'$ (the isometry is understood to act in some neighbourhood of
$H$);
  \item  $\s\equiv\overline H - H$ is an $(n-2)$-dimensional surface and
  $M\equiv\Mf^n-\s$ is non-simply connected with the fundamental group
  $\pi_1=\cel$ (so, in the three-dimensional case $H$ can be, for
  example, a half-plane or a disk but not an infinite cylinder; in
  example~\ref{ex:con}  $H$ is the half-plane $\phi=const$). The
  universal covering of $M$ will be denoted by $\hat M$ and  the
 natural projection $\hat M\to M$  by $\piup$.
\end{enumerate}
Take $V$ to be  $\Mf^n-\overline H$. The spacetime $V$ is extendible and
we shall consider it (not as a spacetime in itself, or a part of $M$,
but) as a part of  $\hat M$ and to indicate this the coordinates in $V$
will be labeled with hats. The boundary of $V$ in $\hat M$ is two
disjoint copies of $H$, which we denote by  $\mathcal B_1$ and $\mathcal
B_2$.

Further, $\pi_1(M)$ is  generated by a single element --- the homotopy
class  of a curve $\ell$ which circles $\eu F$ once. Thus an isometry
$\chiup$ acting on $\hat M$ is defined by the conditions that for any
$x\in\hat M$, first, $\piup(x)=\piup(\chiup(x))$ and, second, there is a
path $\alpha$ from $x$ to $\chiup(x)$
  such that $\piup(\alpha)$ is a loop homotopic to $\ell$
(in example~\ref{ex:con}, $\chiup$ is the translation
$\hat\phi\mapsto\hat\phi+2\pi$). Evidently, $\mathcal B_2=\chiup(\mathcal
B_1)$ and we construct the desired spacetime $\eu{M}$ from
$\overline{\strut V}$ by gluing $\mathcal B_1$ and $\mathcal B_2$ with
the isometry
\begin{equation*}
\kappaup=\kappaup'\circ\chiup
\end{equation*}
(gluing them with $\kappaup=\chiup$ we would get merely $M$).
 In other words, we scissor $H$ from $\Mf^n$, move in a special way one
 bank of
the cut with respect to the other (the motion is devised so as to keep
the boundary of the cut, i.~e.\ $\s$, at place) and glue the banks
together again. The spacetime $\hat M$ was introduced in this procedure
only for giving a rigorous sense to the notion of banks.
\subsection{Examples}\label{sec:spstr}
Let $H$  be the half plane $(\phi=0$, $\rho>0)$ in the three-dimensional
Minkowski space. Then $\s$ is the $z$-axis, $\hat M$ is the spacetime
\[
\rmd s^2= -\rmd \hat t^2 + \rmd \hat \rho^2 + \hat \rho^2\rmd
\hat \phi^2,\qquad
\hat t,\hat \phi\in \rea^1,\quad \hat \rho> 0,
\]
and $V$ is the region $0<\hat\phi<2\pi$  bounded by the half planes
$\mathcal B_1$ and $\mathcal B_2$, which are defined by the equations
$\hat\phi=0$ and $\hat\phi=2\pi$, respectively. If one glued $\mathcal
B_1$ to $\mathcal B_2$ with $\chiup\colon\quad
\hat\phi\mapsto
\hat\phi+2\pi,
$
one would just restore the initial Minkowski space. But if the same
surfaces are glued with the isometry
\[
\kappaup\colon\qquad \quad \hat\phi\mapsto
\hat\phi+2\pi,\quad\hat t\mapsto
\hat t+\hat t_0
\]
(i.~e., $\kappaup=\kappaup'\circ\chiup$, where $\kappaup'$ is the
translation by $\hat t_0$ it the $\hat t$-direction) the result
\cite{3} is a singular spacetime $\eu{M}^-_{4'}$. Its four-dimensional
counterpart $\eu{M}^-_{4}=\eucl^1\times\eu{M}^-_{4'}$ discovered in
\cite{10.,Mazur} is often called `the spinning string'.
Topologically $\eu{M}^-_{4}$ is equivalent to the straight string
$\eu{M}_{1}$. They both are everywhere flat and their singularities are
both represented by flat planes. Nevertheless the two spacetimes differ
significantly in some respects. For one, in $\eu{M}^-_{4}$ chronology is
violated. Another difference is that a tetrad parallel transported along
$\ell$ returns rotated in $\eu{M}_{1}$ but not so in $\eu{M}^-_{4}$. As
we discuss below, there is also a more local difference.

One also can repeat the procedure just described, starting this time from
$\eucl^3$ instead of $\mink^3$ and, correspondingly, shifting
 $\mathcal B_1$ --- before it is glued to  $\mathcal B_2$ --- in the
$z$-direction instead of the $t$-direction
\begin{equation}\label{eq:scrDisl}
\kappaup\colon\qquad \quad \hat\phi\mapsto
\hat\phi+2\pi,\quad\hat z\mapsto
\hat z+\hat z_0.
\end{equation}
In this case the result is the spacetimes $\eu{M}^+_{4'}$  and
$\eu{M}^+_{4}$ called \emph{screw dislocations} for their similarity to
the corresponding distortion \cite{GL}.
 $\eu{M}^+_{4}$ differs both from $\eu{M}_{1}$ and $\eu{M}^-_{4}$ though its
singularity is also represented by a plane.

Finally, one can start  from $\mink^4$ (in this case $\mathcal B_j$ are
given by the same equations, but now they are three-dimensional half
spaces) and choose $\kappaup$ to be a combination of $\chiup$ with a
boost in the $z$-direction
\cite{Tod}. Thus obtained spacetime $\eu{M}_{5}$ has a number of curious
properties. For example, it is not even stationary, even though its every
simply connected region is static.

Sometimes string-like singularities can be built by `superposing'
elementary ones. Take, for example, $V$ to be the sector $0<\hat\phi<a$
in $\hat M$ and $\kappaup$ to be the superposition of translations: by
$a$ in the $\hat\phi$-direction, by $\hat t_0$ in the $\hat t$-direction,
and by $\hat z_0$ in the $\hat z$-direction. The result is the spacetime
\cite{GL,Tod}
\begin{equation}\label{eq:GLT}
\begin{split}
 \rmd s^2= -(\rmd \hat t-a^{-1}\hat t_0\,\rmd\hat \phi)^2 + (\rmd \hat z^2
 -a^{-1}\hat z_0\,\rmd\hat \phi)^2
 +\rmd \hat \rho^2+  \hat \rho^2\rmd\hat \phi^2,
\\ \hat \phi=\hat \phi+a.
\end{split}
\end{equation}
which combines the properties of the three spacetimes.
\subsection{The strength of the singularities.}
In discussing singular spacetimes it is often hard to decide whether a
particular property should be considered as a characteristic of the
singularity or of the `regular part' of the spacetime. For example, it
seems natural to classify the string-like singularities according to
their holonomies \cite{PS,Tod}. On the other hand,
example~\ref{ex:branch} below suggests that such a classification may be
misleading. Fortunately, the simplicity of the spacetimes at hand, allows
a quantity to be found which seemingly relates just to the singularity.
The cost is that the corresponding classification is quite rough --- the
singularities are divided  only into three categories. One of them
contains the singularities from appendix~\ref{A:ES} and the other
contains  those from section~\ref{sec:spstr}.

Let $\eu{s}_\gamma$ be a singular point, $\gamma(\tau)$ be a geodesic
defining this point as was discussed above, and $\{\boldsymbol e_{(i)}\}$
be an orthonormal frame in  $\gamma(-1)$. Now if a curve
$\alpha(\xi)\colon [\xi_1,\xi_2]\to M$ starts from some point of $\gamma$
\[
\alpha(\xi_1)=\gamma(\tau_1),
\]
it is possible to assign to it the `b-length' $B(\alpha)$ . To this end
one defines $\{\boldsymbol e_{(i)}\}(\xi)$ to be the frame in
$\alpha(\xi)$ obtained by parallel transportation of $\{\boldsymbol
e_{(i)}\}$, first, along $\gamma$ to the point $\gamma(\tau_1)$ and then
along $\alpha$. Then $B(\alpha)$ (it is the length of $\alpha$ in the
generalized affine parameter, see
\cite{HawEl}) is defined as
follows:
\[
B(\alpha)=\int_{\xi_1}^{\xi_2}
\Big[\sum_i
\langle\partial_\xi,\boldsymbol e_{(i)}\rangle^2\Big]^{1/2}
\,\rmd\xi.
\]
Now we can attach a number
\[
\Delta(\eu{s}_\gamma)\equiv\lim_{\tau\to 0}
\inf_{\substack{\text{noncontractible}\\ \text{$\alpha$ through $\gamma(\tau)$}}}
B(\alpha)\]
 to every point $\eu{s}_\gamma$. Of course the value of
$\Delta(\eu{s}_\gamma)$ for a given $\eu{s}_\gamma$ may depend on
$\gamma$ and $\{\boldsymbol e_{(i)}\}$, but not when
\[
\Delta(\eu{s}_\gamma)=0\tag{$*$}.
\]
The property ($*$) holds for \emph{all} (equivalent) $\gamma$ and all
$\{\boldsymbol e_{(i)}\}$ if it holds for some.

The spacetimes in which  ($*$) is true for all $\eu{s}\in\s$ will be
called \emph{disclinations} after the spacetime  $\eu{M}_{1}$,  which is
often called so (by analogy with the theory of elasticity). In fact, all
spacetimes constructed in example~\ref{ex:con} and appendix~\ref{A:ES}
are disclinations. On the contrary, the spacetimes $\eu{M}_{4}^\pm$,
$\eu{M}_{5}$ all are characterized by the opposite property: ($*$) holds
in none of $\eu{s}\in\s$. I shall use the word
\emph{dislocation}\footnote{At  variance with Puntigam and Soleng, who
divided the spacetimes into disclinations and dislocations according to
their \emph{global} properties \cite{PS}.} as a common name for all such
spacetimes.

Absolutely mild singularities are often referred to as `topological'.
However, when it concerns  the disclinations such a name may be a bit
misleading. What makes these singularities `true' (i.~e.\ irremovable) is
the purely
\emph{geometrical} requirement that the metric should be non-degenerate.
In a hypothetical theory in which this requirement is
relaxed\footnote{Such a theory would differ significantly from general
relativity.} there would be no singularity at all except maybe a
`coordinate singularity' like that on the horizon of the Schwarzschild
black hole or in the origin of the polar coordinates. Indeed, the
spacetime~\eqref{eq:cone_Eu}, for instance, can be extended to
$\rea^2\times\okr^1$ by simply letting $\rho$ vary over the entire real
axis. The only pathology is that $g=0$ at $\rho=0$.

Dislocations in this sense are stronger singularities. As is seen from
the definition, $\s$ cannot be be retained in the spacetime even at the
cost of the metric degeneration (as long, that is, as only continuous
metrics are allowed).
\section{Unconventional singularities}\label{Sec:non-ES}

In this section a few string-like singularities are constructed with
rather unusual properties. To my knowledge none of them, except
$\eu{M}_{10}^\pm$, have been considered in the literature.
\subsection{Curved, closed, and accelerating dislocations}
In the Euclidean space $\Mf^3=\eucl^3$ consider the surface $H$
\[
 \phi=bz\mod2\pi, \qquad\rho>\rho_0>0,\qquad\quad b\neq 0.
\]
$H$ is a half of a helicoid without the core, see
figure~\ref{fig:spiral},
\kartih{tb}{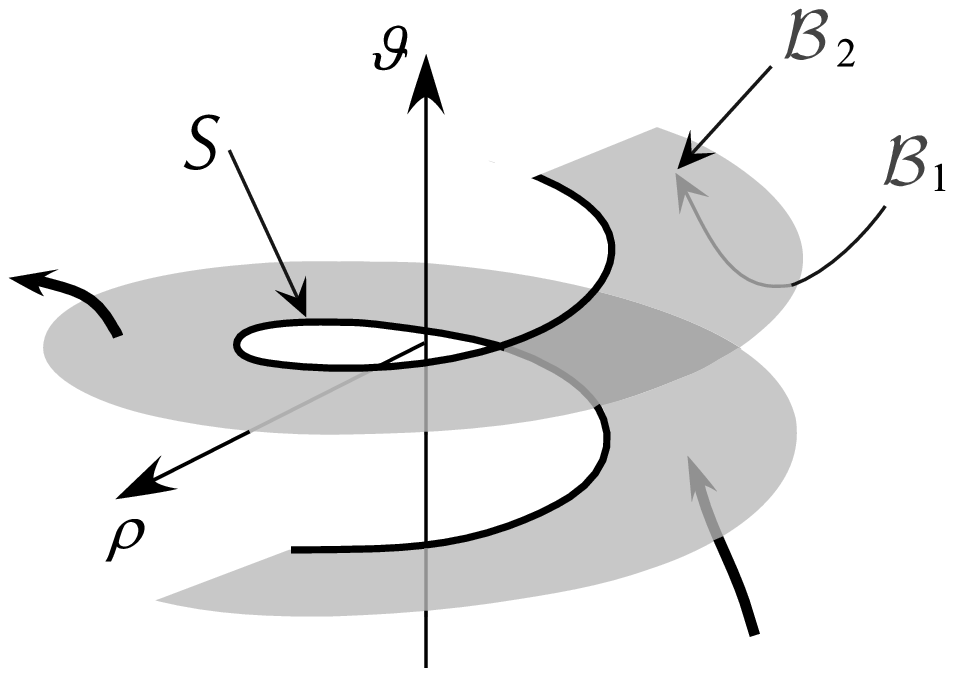}{\label{fig:spiral} The spacetime $\eu{M}^+_{6'}$. The
thick directed line is continuous.} and is bounded by the spiral
\[
\s\colon\ \quad \phi=bz\mod2\pi,
\quad \rho= \rho_0.
\]
With these $\Mf^3$ and $H$ let us carry out the procedure described in
the end of section~\ref{sec:surg}. The boundary of $\Mf^3-H$ in $\hat M$
consists of  $\mathcal B_1$ and $\mathcal B_2=\chiup(\mathcal B_1)$ (two
disjoint copies of $H$) and there is an obvious isometry
\begin{equation}\label{eq:spir}
\kappaup'\colon\qquad
\hat\phi\mapsto
\hat\phi+ b\hat z_0,\quad\hat z\mapsto
\hat z+ \hat z_0
\end{equation}
that maps $\mathcal B_2$ to itself.  So, gluing it to $\mathcal B_1$ with
$\kappaup=\kappaup'\circ\chiup$ we obtain a spacetime $\eu{M}^+_{6'}$
with a singularity represented by the spiral $\s$. (In other words, we
have made a cut in  $\eucl^3$ along the helical surface, rotated the
lower bank of the slit --- let it be $\mathcal B_1$ for definiteness ---
counterclockwise shifting it at the same time upward so that $\mathcal
B_1$ slides over $\mathcal B_2$, and pasted the banks together again into
a single surface.) The singularity $\s$ satisfies the relation
\[
\Delta(\eu{s})=\sqrt{(2\rho_0\sin\tfrac12 b\hat z_0)^2
+\hat z_0^2}\qquad\forall\,\eu{s}\in\s,
\]
being thus a dislocation. I shall call it \emph{spiral} (not to be
confused with helical).

To realize the structure of the spiral dislocation it is instructive to
depict $\eu{M}^+_{6'}$ in the coordinates $z$, $\rho$, $\phi$ as in
figure~\ref{fig:spiral}. These coordinates are invalid, of course, on
$\mathcal B_{1,2}$; that is why a smooth curve in $\eu{M}^+_{6'}$ looks
discontinuous in the picture. It is easy to see that at $b\hat z_0
\neq 2\pi$ the geometry of the space outside the cylinder $\rho\leq\rho_0$ is
exactly the same as in (the three-dimensional version of) the
Gal'tsov-Letelier (GL) space\footnote{The difference in presentation
yields an interesting by-product: the Gal'tsov-Letelier space is not
defined at $a=0$, while $\eu{M}^+_{6'}$ is a nice spacetime for any $\hat
z_0$}
\eqref{eq:GLT} with  $a=b\hat z_0 - 2\pi$. The relation of the two
spacetimes becomes even more evident when the spacetime is considered
which is obtained exactly as $\eu{M}^+_{6'}$ but with the surface $H'$
\[
H'\colon\ \quad  \phi=bz\mod2\pi, \qquad0<\rho< \rho_0.
\]
taken instead of $H$.
\kartih{tb!}{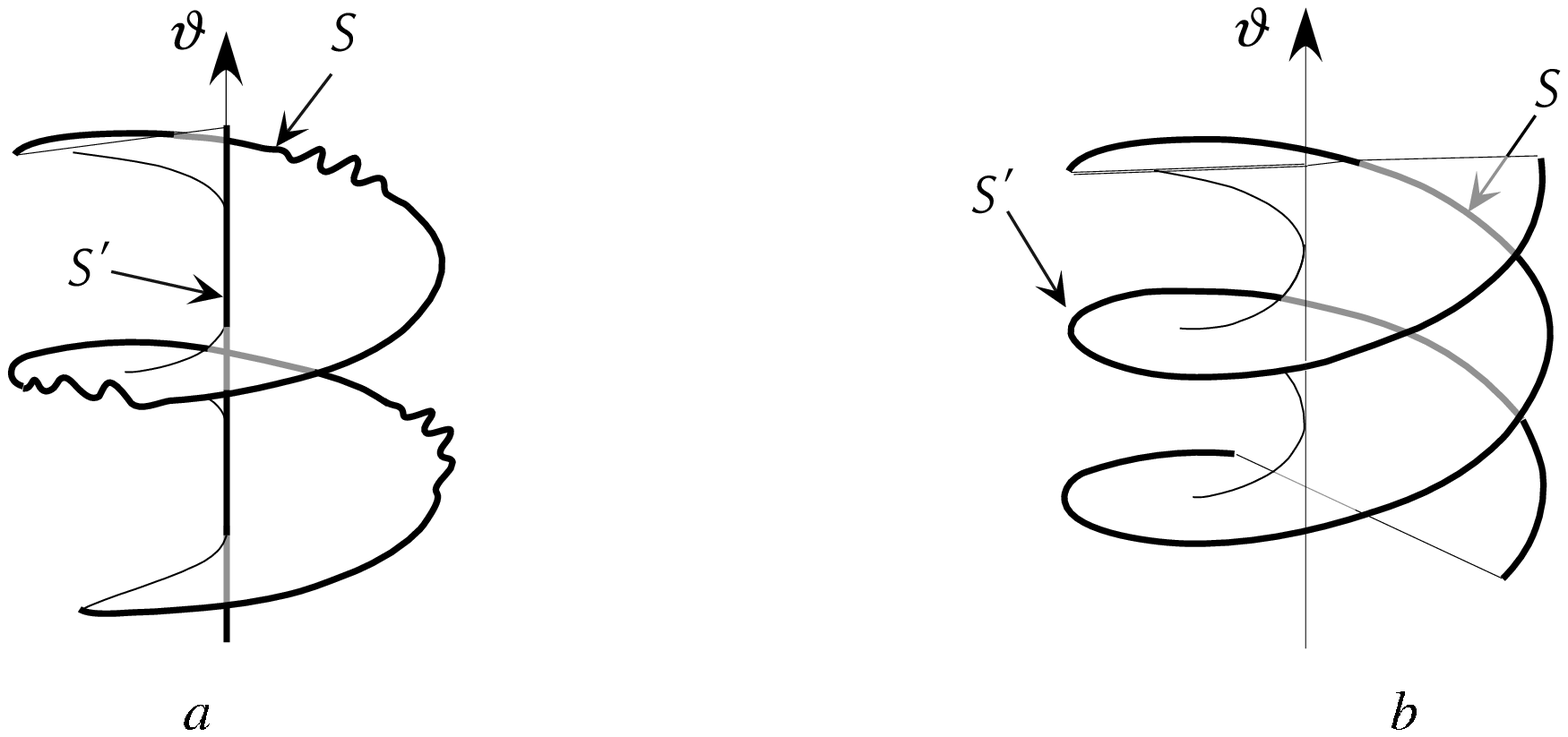}{\label{fig:helicoid} a). A spiral singularity
shielding a GL one. b). Two equal spiral singularities shielding each
other.}%
In contrast to  $H$, $H'$ is bounded by $\s$
\emph{and} the $z$-axis. Thus the spacetime (see figure~\ref{fig:helicoid}a ignoring for the
moment the `ripples' on $\s$) has two singularities, of which $\s$ is
spiral and $\s'$ (the former $z$-axis) is of the GL type. At $\rho>
\rho_0$ the spacetime is just $\eucl^3$, so the spiral singularity
\emph{`shields'} the GL one. And in exactly the same sense two equal spiral
singularities can shield each other. The spacetime of this type is built
by taking $H$ to be the central part of a helicoid, see
figure~\ref{fig:helicoid}b.

The most striking feature of $\eu{M}^+_{6'}$ is of course the form of the
singularity. From all the preceding examples it might seem that
string-like singularities in flat spacetime by some reason have to be
straight. And now we see that this is not the case
--- they may well be curved. Note that $\s$ does
not need to be a perfect spiral --- instead of $H$ we could take another
surface as long as it is invariant under the isometry \eqref{eq:spir} and
its boundary, exemplified by the undulate line in
figure~\ref{fig:helicoid}a,
\kartih{tb}{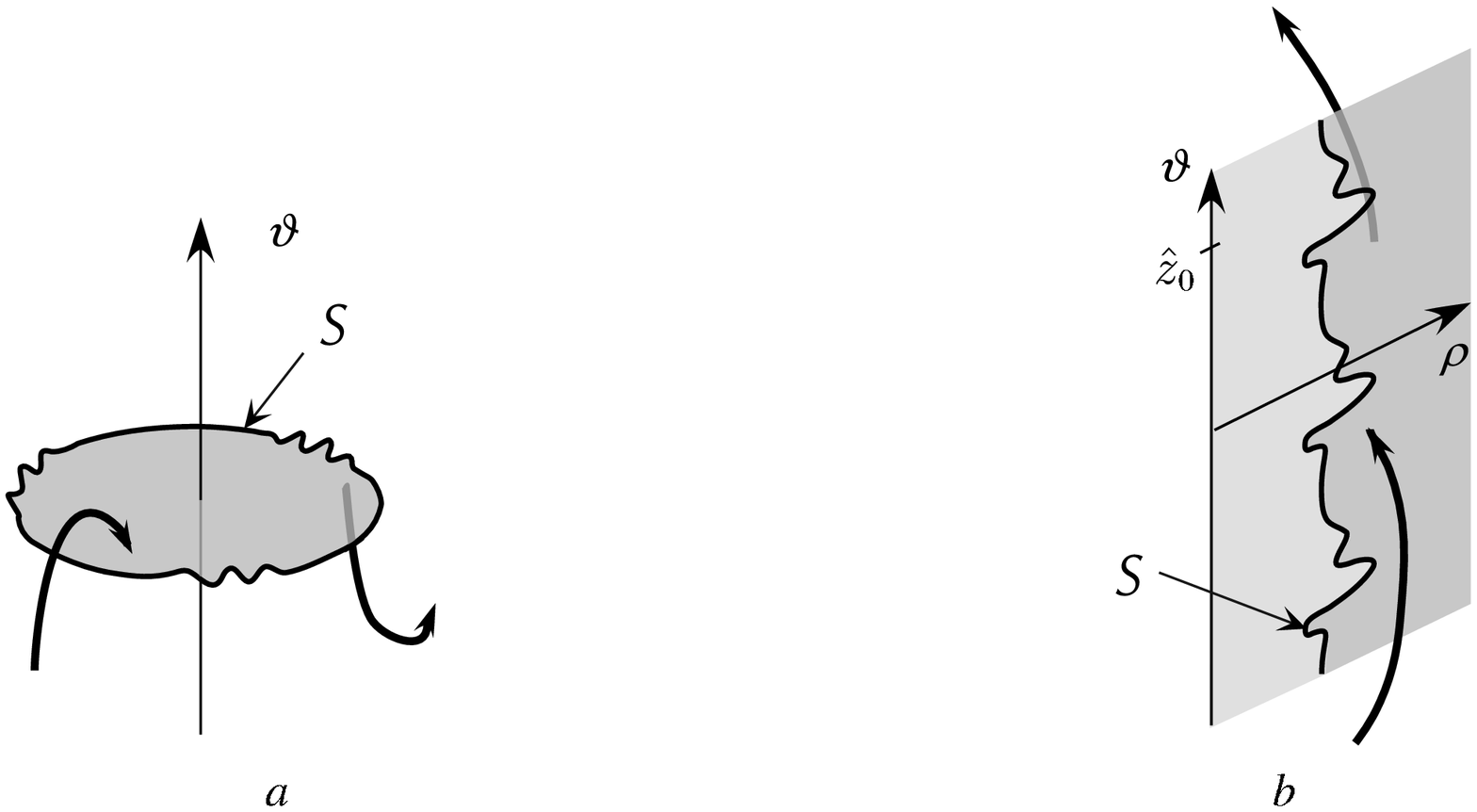}{\label{fig:pentapet} a). A loop singularity with
$\phi_0=\tfrac23\pi$. b). Curved screw dislocation.} will represent a
singularity of exactly the same type. Moreover, $\s$ can even make a
\emph{loop}. Indeed, pick a closed curve $\s\subset
\eucl^3$ bounding a surface $H$ invariant under the rotation
\[
\kappaup'\colon\qquad
\phi\mapsto
\phi+ \phi_0.
\]
($H$ and $\s$ needn't be a surface of revolution and a circle,
respectively, if $\phi_0= 2\pi/m,\ m\in
\nat$). Proceeding as before (i.~e.\ making an incision along $\overline
H$ and gluing the banks together after rotating one of them by $\phi_0$)
we obtain a space $\eu{M}^+_{7'}$ with a
\emph{closed} string-like singularity, see figure~\ref{fig:pentapet}a.
$\eu{M}^+_{7'}$ can be viewed as a limit case of the spiral singularity
corresponding to $b=\infty$. Another limit, $b=\rho_0=0$ is the space
$\eu{M}^+_{8'}$ depicted in figure~\ref{fig:pentapet}b. It is a pure
screw dislocation, but
\emph{curved}.  This space is built exactly as  $\eu{M}^+_{5}$
with the only difference that instead of the entire half-plane $\phi=0$
one takes $H$ to be the part of that  half-plane lying above the graph of
 a periodic function $\rho(z)$
\[\rho(z)\geq 0,\qquad \rho(z)=\rho(z+\hat z_0).
\]

A different family of dislocations, $\eu{M}^-_{6',7',8'}$, is obtained
when the surgery employed in constructing $\eu{M}^+_{6',7',8'}$ is
applied to $\mink^3$ instead of $\eucl^3$. Correspondingly, their
four-dimensional versions $\eu{M}^-_{6,7,8}$ are obtained by
interchanging the $z$- and $t$-axes in $\eu{M}_{6,7,8}^+$. Of these
especially interesting is $\eu{M}_{6}^-$, see figure~\ref{fig:spiral}
with $\vartheta=t$. At $\rho>\rho_0$ it is just a spinning string, but
taken as a whole it has two important distinctions. First, $\eu{M}_{6}^-$
in a large range of its parameters $b$, $\hat z_0$, and $\rho_0$ is
\emph{causal}. And, second, the singularity there is represented by a
straight line moving  in quite a bizarre manner: it \emph{circles around
nothing}.

Yet another accelerated singularity
--- let us call it $\eu{M}_{9}$ --- results from choosing
$H$ to be the 3-space $x>\sqrt{t^2+c^2},\ y=0$ in $M=\mink^4$ and
$\kappaup'$ to be a boost in the $x$-direction. The singularity in
$\eu{M}_{9}$ is represented by a straight line parallel to the $z$-axis
and moving with a constant acceleration in the $x$-direction. If such a
string passes between two observers, which initially are at rest
w.~r.~t.\ each other, either of them would suddenly discover that the
other has acquired some speed in the $x$-direction, even though no
apparent forces were involved.

\subsection{Curved disclination}\label{sec:free}
In an arbitrary spacetime $M$ pick a surface $\s$ of co-dimension 2 such
that curves wrapping around it are non-contractible in $M-\s$.  Consider
the $i$-fold covering of $M-\s$. Irrespective of what $M$, $\s$ and $i$
are chosen ($i$ must be finite, though) the covering has a  string-like
singularity represented by $\s$. It is easy to see that this `branching
singularity' is a disclination corresponding to $\zetaup=\mathop{\rm id}$
in terms of appendix~\ref{A:ES}. As such, the singularities of this type
have received surprisingly little attention in the literature, however,
implicitly they are present in a number of known spacetimes.

Let $\kappaup$ be an isometry sending an open subset $O_1$ of a spacetime
$M$ to a subset $O_2$ disjoint with $O_1$.
\kartih{tb}{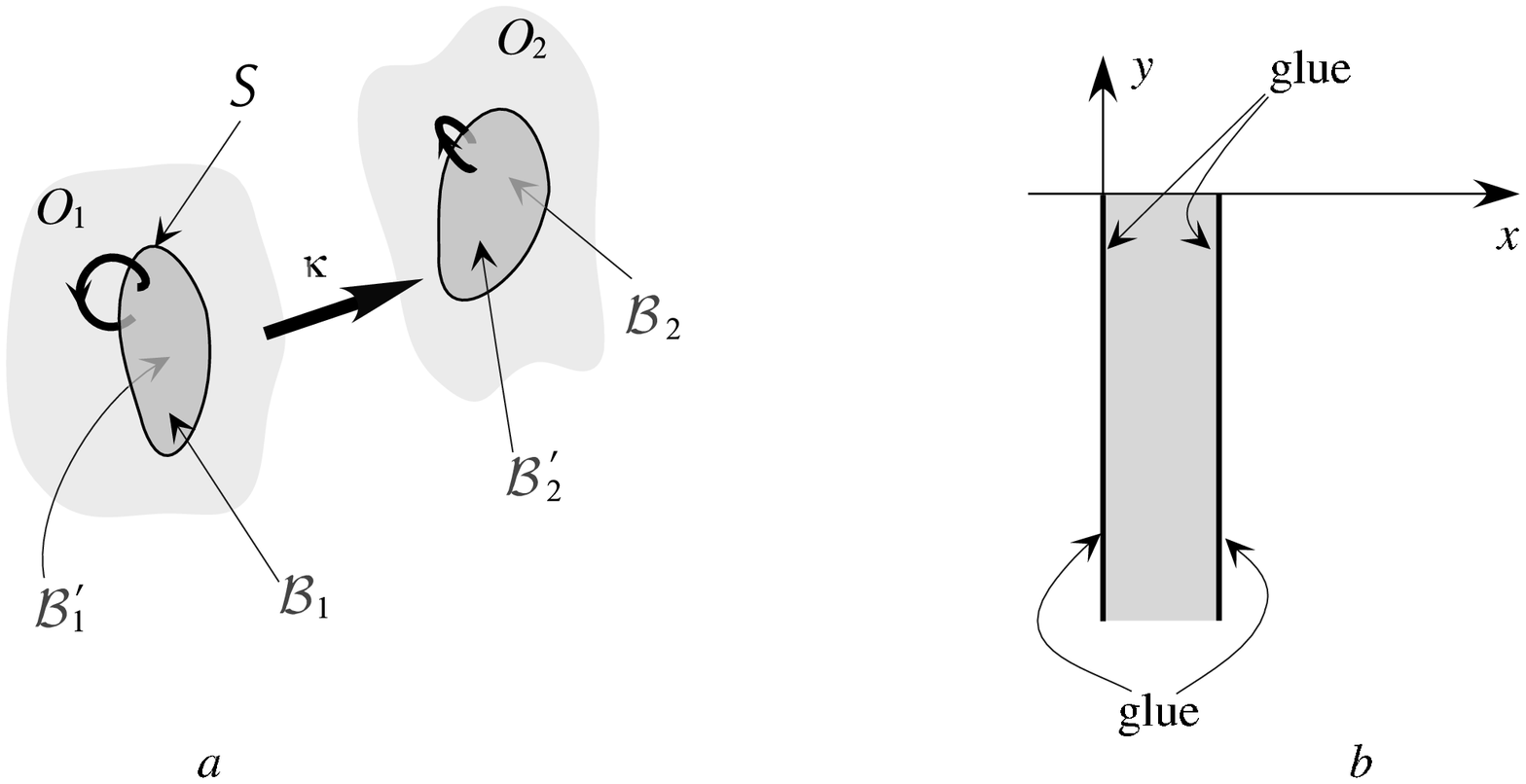}{\label{fig:free}a). $\mathcal B_1$ is glued to
$\mathcal B_2$ and $\mathcal B'_1$ to $\mathcal B'_2$. The thick curve
is, in fact, continuous (and closed). b). The thick vertical lines are
slits.}
To construct a new spacetime $M'$ pick a two-sided hypersurface $H$ lying
in $ O_1$ together with its closure $\overline {H}$, remove $H$ and
$\kappaup(H)$ from $M$ and glue the corresponding banks of the two thus
obtained slits as is shown in figure~\ref{fig:free}a. Then $M'$ has a
singularity represented by $\s=\overline {H} - H$ which is just a
branching dislocation\footnote{To give a precise meaning to the word
`banks' and to make the procedure rigorous in every way one should
consider $M- \overline {H}-\overline {\kappaup(H)}$ as a region in a
covering of $M-\s-
\kappaup(\s)$ and proceed as in section~\ref{sec:surg}.}
--- the gray region in the figure being nothing but the double covering
of $O_1-\s$.

\begin{teor}{Example}\label{ex:branch}
When $H$ is a spacelike disc in the Minkowski space and $\kappaup$ is a
timelike translation combined with the time reflection,
$\eu{M}_{10}^-=M'$ is the Deutsch-Politzer space. It contains closed
causal curves and due to its simplicity is used extensively in  time
machine theory \cite{Tm}. Its Euclidean analog $\eu{M}_{10}^+$ is a
`loop-based wormhole'  constructed (in terms of `delta function Riemann
tensor') in~\cite{Viss}. For its thickened version see~\cite{portal}
(curiously,
 in $\eu{M}_{10}^-$ the singularity can\emph{not} be thickened
\cite{nosmDP}).
\end{teor}

If nothing else, the branching singularity is a wonderful source of
counter-examples. It shows, in particular, that in the \emph{general}
case
\begin{enumerate}
  \item The presence of a string-like singularity puts \emph{no}
  restrictions on  the stress-energy tensor of the hosting spacetime.
  \item A string-like singularity can take any form and change it
  arbitrarily (though in a smooth manner, of course) with time. It also can appear and
  disappear at will (for example, $\s$ can be a circle in the
  $(t,x)$-plane multiplied by the $z$-axis). So, there are
  no `laws of motion' for a general singularity (and, in particular, it
  does not have to be a plane, contrary to what is asserted in \cite{Vickers}).
  \item Holonomies do not characterize string-like singularities, even
  disclinations. Indeed, in
  constructing $\eu{M}_{10}$ we could vary $\kappaup$ (combining, say,
  the translation with a rotation) and obtain spacetimes
  with different holonomies even though the singularities remain the
  same.
\end{enumerate}
Incidentally,  the first two facts mean that the Cosmic Censorship
Conjecture can be proved \emph{only} if general relativity is
complemented by an additional global postulate, like hole-freeness
(cf.~\cite{quasireg}).

\subsection{Mixed  singularities}\label{sec:mixed}

Remove from the Euclidean plane $\eucl^2$ the region
\[
0<x<1, \qquad y\leq 0
\]
(the gray strip in figure~\ref{fig:free}b) and glue its vertical
boundaries:
\[
(0,y)\mapsto (1,y)
\]
to obtain a new 2D spacetime $\eu{M}_{e}$. One might think that
$\eu{M}_{e}^4=\mink^2\times\eu{M}_{e}$ is a spacetime with a string-like
singularity of yet another type, which could have been called `edge
dislocation' for its similarity to the corresponding distortion in solid
state science (and that is, indeed, how $\eu{M}_{e}^4$ --- or, rather,
its part --- was called in
\cite{PS}, see Appendix~\ref{A:PS}).
In fact, however, this is not the case, because $\eu{M}_{e}$ is
\emph{extendible} and hence its singularities, formally speaking, are not even
absolutely mild. It is easy to find an inextendible extension of
$\eu{M}_{e}$: such is for example, the spacetime $\eu{M}_e^\text{ext}$
obtained by making a pair of slits
\[
x=0,1, \qquad y\leq 0
\]
in  $\eucl^2$ and gluing their banks as is shown in
figure~\ref{fig:free}b. It is seen that there \emph{is} a singularity in
$\eu{M}_e^\text{ext}$, but this is just a branching singularity discussed
above (the only difference between $\eu{M}_e^\text{ext}$ and a 2D
loop-based wormhole is that the slits are semi-infinite in the former and
finite in the latter).

The interrelation between the edge dislocation and the branching
singularity suggests (again by analogy with the solid state physics) that
the latter can `transform' into a screw dislocation forming thus a
\emph{mixed} singularity. And this is the case.
\kartih{htb}{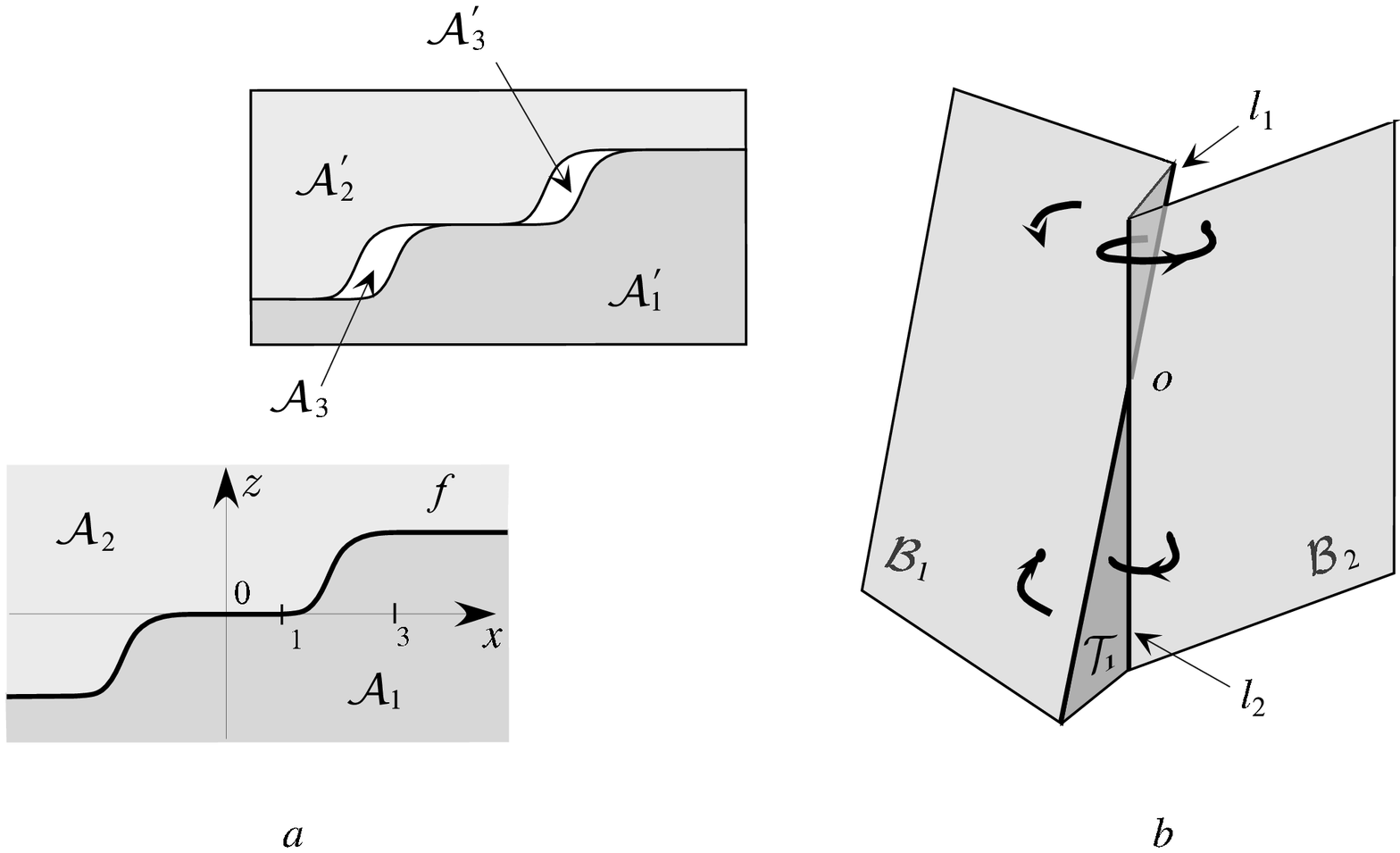}{\label{fig:str} a).  Each $\mathcal A_{m}$ is
glued to $\mathcal A'_{m}$. b).  $V'$ is the part of the space separated
from us by $\mathcal B$. The curves with arrows make in fact a single
smooth path.} To build an example remove the  plane $y=0$ from $\eucl^3$
and attach two banks
--- each is a copy of the  plane
--- to the cut as explained in section~\ref{sec:surg}. On one of the
banks draw a curve $f$ which is the graph of a smooth monotone function
$z(x)$:
\[
z=
  \begin{cases}
    -1 & x<-3, \\
    0 & |x|<1,\\
    1& x>3.
  \end{cases}
\qquad
z(x)=1+z(x-3) \quad\text{at } x\in (0,3)
\] see figure~\ref{fig:str}a. This
curve splits the bank into two regions which I denote  $\mathcal A_{1}$
and $\mathcal A_{2}$. On the other bank draw two lines --- one is $f$ and
the other is obtained from it by a horizontal shift $x\to x+\frac12$. The
bank thus is split into four regions denoted $\mathcal A'_{1}$, $\mathcal
A'_{2}$, $\mathcal A_{3}$ and $\mathcal A'_{3}$. To assemble the
spacetime $\eu{M}^+_{11'}$ remove all three copies of $f$ and paste each
$\mathcal A_{m}$ to the corresponding $\mathcal A'_{m}$, $m=1,2,3$
($\kappaup$ for $m=3$ is a combination of the shift $(x,z)\mapsto (x+3,
z+1)$ and reflection  $y\mapsto -y$). As can be easily seen,
$\eu{M}^+_{11}$ has a string-like singularity represented by $f$. The
singularity is mixed: in particular, it is a (branching) disclination at
$x=2$ and a screw dislocation at $x=4$.

The last example is built as follows. In $\eucl^3$ pick two intersecting
straight lines $l_1$ and $l_2$ and attach a half-plane to either of them
as shown in figure~\ref{fig:str}b. Together these half-planes $\mathcal
B_{1,2}$ and the angles $\mathcal T_{1,2}$ bounded by the lines $l_{1,2}$
form a surface $\mathcal B$ that divides the space into two disjoint
regions $V$ and $V'$. There are two obvious rotations, one of which
(denoted $\kappaup_1$)  maps $\mathcal B_{1}$ to $\mathcal B_{2}$ and the
other --- $\kappaup_2$ --- maps $\mathcal T_{1}$ to $\mathcal T_{2}$. The
spacetime $\eu{M}_{12'}$ is obtained by throwing away  $V'$ and pasting
$\mathcal B_{1}$ to $\mathcal B_{2}$ with $\kappaup_1$ and $\mathcal
T_{1}$ to $\mathcal T_{2}$ with $\kappaup_2$. The singularity in
$\eu{M}_{12}$ is represented by a plane $l_1\times t$-axis and yet it is
none of the singularities considered in section~\ref{sec:gen}. In
particular, it is neither a disclination nor a dislocation, since
$\Delta(\eu{s})=0$ at $\eu{s}=l_1\cap l_2$ and $\neq 0$ otherwise.
\section*{Acknowledgements}
This  work was  supported by RNP Grant No.~2.1.1.6826.

\appendix
\section{Appendix}\label{A:ES}
Denote by $\eu F$ a straight line or a plane (depending on whether $n=3$
or 4) which is the set of fixed points of an isometry $\zetaup\colon\
\Mf^n\to
\Mf^n$:
\begin{equation*}
 \eu F= \{p\in \Mf^n:\quad\zetaup(p) = p\}.
\end{equation*}
Let  $\hat M$ be the universal covering of  $M\equiv\Mf^n-\eu F$. Define
\[
\hat\zetaup_k\colon\qquad \hat M\to\hat M
\]
to be a lift of $\zetaup\circ\piup$ (here $k\in\cel$, since every point
of the fiber defines a new lift). If, for example, $\zetaup$ is the
rotation by $a$ in $\eucl^3$, then $\hat M$ is given by
\eqref{eq:Mcov} and $\hat\zetaup_k\colon\
\hat\phi\mapsto\hat\phi+ a+2\pi k$.

It is the quotients $\eu{M}=\hat M/\hat\zetaup_k$ that Ellis and Schmidt
tested for absolutely mild singularities. Indeed, the construction of
$\eu{M}$ is a generalization of that producing a cone or the Misner
space, so it is reasonable to expect (though not
\emph{guaranteed}) that  $\eu{M}$ has a string-like singularity
represented by $\eu F$.

One spacetime of that kind is $\eu{M}_{1'}^+$, which already has been
constructed in example~\ref{ex:con}. The four-dimensional spacetime
$\eu{M}_{1}$ with a string-like singularity (in this case
$\eu{M}_{1}=\eu{M}^+_{1}=\eu{M}^-_{1}$) is obtained as the product
$\eu{M}_{1'}^+\times\mink^1$. Two more spaces of this type were found in
\cite{quasireg}; let us denote them $\eu{M}_{2}$ and $\eu{M}_{3}$. The
spacetime $\eu{M}_{2}$ is obtained by taking $\eu F$ to be a spacelike
surface in the Minkowski space and $\zetaup$ to be a boost in the
direction perpendicular to  $\eu F$. [Interestingly enough not
\emph{all} $k$ are equally appropriate in this case:  one particular
$k_0$ (that for which $\hat M/\hat\zetaup_{k_0}=M/\zetaup $)  must be
excluded, because the quotient is non-Hausdorff.] Finally, $\eu{M}_{3}$
is built exactly as $\eu{M}_{2}$ but with $\eu F$ being null and
$\zetaup$ being, correspondingly, the combination boost+rotation which
leaves all points of $\eu F$ fixed. Clearly,  all three spacetimes
$\eu{M}_{1,2,3}$ have string-like singularities represented by the planes
$\eu F$.

\section{Appendix}\label{A:PS}

In their paper \cite{PS} Puntigam and Soleng employed  the Volterra
process to obtain flat spacetimes with unusual holonomies and thus with
singularities. Two of them (see entries 1 and 2 of table 2) are called
`edge dislocation'. The goal of this Appendix is to demonstrate that
these two spacetimes (they are isometric) are, in fact, regions in the
spacetime $\mink^2\times\eu{M}_e$ considered in the beginning of
section~\ref{sec:mixed}.

The spacetimes in discussion are $\mink^2\times\eu{M}_{\text{PS}}$, where
the metric of $\eu{M}_{\text{PS}}$ is
\[
\begin{split}
\rmd s^2 =  \rmd x'^2 + \rmd y^2
          + 2\frac{\Theta^1}{2 \pi r^2} \rmd x' (x' \rmd y - y \rmd x')
          + \left(\frac{\Theta^1}{2\pi r^2}\right)^2(x' \rmd y - y \rmd
          x')^2,\\r^2\equiv x'^2+y^2.
\end{split}
\]
To analyze the structure of $\eu{M}_{\text{PS}}$ let us first rewrite the
metric in a more transparent way:
\begin{equation}\label{eq:Mps}
\begin{split}
\rmd s^2=(1+\theta^2x'^2)\rmd y^2 +(1-\theta y)^2\rmd x'^2 +2\theta x'(1-\theta y)
\rmd x'\rmd y\\
\theta\equiv{2\zeta}/{ r^2},\qquad \zeta\equiv\tfrac{1}{4\pi}\Theta^1.
\end{split}\end{equation}
It is easy to see now that the metric  diverges  at $r=0$ and degenerates
at $\theta y=1$, i.~e.\  on the circle
\begin{equation}\label{eq:circle}
(y-\zeta)^2 + x'^2=\zeta^2.
\end{equation}
Its domain consists thus of two \emph{disjoint} regions (since the metric
\emph{must} be non-degenerate). Restricting ourselves to the larger
region (i.~e.\ to the exterior of the circle) we conclude that the
spacetime $\eu{M}_{\text{PS}}$ is the manifold
\[
N=\rea^2 - \{(y-\zeta)^2 + x'^2\leq\zeta^2\}
\]
endowed with the metric~\eqref{eq:Mps}. In the coordinates $x'$, $y$ the
manifold $N$ has the appearance shown by gray in
figure~\ref{fig:degen}a).
\kartih{htb}{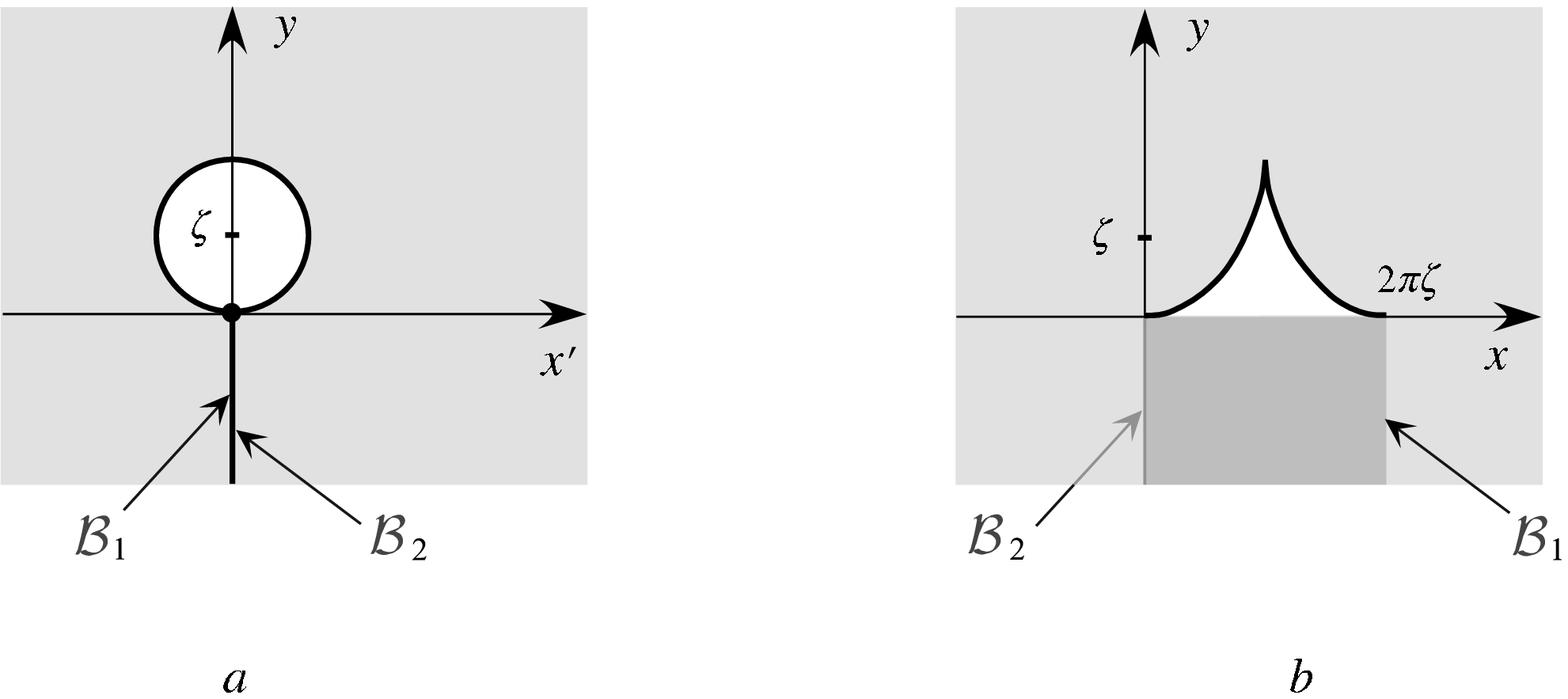}{\label{fig:degen} $\eu{M}_{\text{PS}}$ is the
manifold on the left with the metric \protect{\eqref{eq:Mps}} or,
equivalently, the manifold on the right [the $\curlywedge$-like curve is
the cycloid to which the transformation
\protect{\eqref{eq:cikl}} sends the circle \protect{\eqref{eq:circle}};
the dark gray indicates that we see one sheet through another] with the
metric
\protect{\eqref{eq:standard}}. $\mathcal B_1$
and $\mathcal B_2$ in both cases are glued.}%
It is instructive, however, to introduce a new coordinate $x$
\begin{equation}\label{eq:cikl}
x(x',y)\equiv  x' + {2 }\zeta\arctan y/x',\qquad (x',y)\in N',
\end{equation}
where $N'$ is $N$ without the semi-axis $\{x'=0,\ y<0\}$.
\begin{teor}{Remark}
The cut is necessary to make $\arctan$ well defined, but it is made in
the domain of the function $x$, not in the spacetime being discussed.
\end{teor}
In the coordinates $x$, $y$ the metric takes the form
\begin{equation}\label{eq:standard}
\rmd s^2 =  \rmd y^2
+\rmd x^2,
\end{equation}
while $N'$ becomes the surface shown in figure~\ref{fig:degen}b (and $N$
ensues when $\mathcal B_1$ is glued to $\mathcal B_2$).

\end{document}